\documentstyle[aps,psfig,multicol]{revtex}

\begin{document}
\title{Stability of Solid State Reaction Fronts}
\author{G. Grinstein, Yuhai Tu, and J. Tersoff}
\address{IBM Research Division, T.J. Watson Research Center, P.O. Box 218,\\
Yorktown Heights, NY 10598}
\date{\today }
\maketitle

\begin{abstract}
We analyze the stability of a planar solid-solid interface at which a
chemical reaction occurs. Examples include oxidation, nitridation, or
silicide formation. Using a continuum model, including a general formula for
the stress-dependence of the reaction rate, we show that stress effects can
render a planar interface dynamically unstable with respect to perturbations
of intermediate wavelength.
\end{abstract}

\begin{multicols}{2}
\narrowtext

Many important reactions occur at solid-solid interfaces, and require one of
the reacting species to diffuse to the interface through one of the solids.
The oxidation of silicon, a complex process vital to the fabrication and
function of silicon devices, is the best known example; but the oxidation of
metals, nitridation of silicon, and silicide formation fall into the same
general category. Since the original and reacted materials typically have
different lattice constants, the reaction generates stress, which in turn
alters the reaction rate. In non-planar geometries, where the stress is
non-uniform, the resulting structure can be drastically affected \cite
{Kao,Liu,beak}. There is a longstanding effort to understand these stress
effects on morphology in Si oxidation and in other systems \cite
{Kao,Liu,beak,silicide,Freund,Aziz}.

It is well known that in the case of epitaxial growth of a solid at a free
surface, stress causes a morphological instability in an initially planar
surface\cite{Grin}. In this paper, we show that the stress at a solid-solid
reaction front can similarly lead to an instability \cite{prior}.

The instability here, however, differs in two important respects from that
occurring in epitaxial growth. First, the instability in epitaxial growth is
driven by the thermodynamics, always acting to lower the energy of the
system. In contrast, the reaction instability is essentially dynamical in
nature. It results from the effect of stress on the {\it reaction rate}
rather than on energetics. Stress may either stabilize or destabilize the
planar reaction front, even though the energy is always lowered by
long-wavelength deviations from planarity. Second, a free surface under
stress is always unstable at long wavelengths and stable at short
wavelengths. In contrast, solid-solid reaction fronts are stabilized at long
and short wavelengths by diffusion and interface-tension effects,
respectively. An instability can therefore occur only at intermediate
wavelengths.

The system we study is illustrated schematically in Fig.\ 1. For simplicity,
we use the language of oxidation to describe it. However, the model we now
discuss is quite general, and applies equally well to the other reactions
mentioned above. In oxidation, the surface layer of a solid (typically metal
or semiconductor) is in contact with a reservoir of oxidant, such as $O_2$
gas. The surface oxidizes, producing a thin solid film of oxide, as shown in
Fig.\ 1. The oxidant must diffuse through the film to the film-substrate
interface in order for further oxidation and film growth to occur.

Typically there is some volume change upon oxidation, which produces stress
\cite{Kao}. This in turn affects the oxidation and diffusion rates. In some
materials, such as SiO$_{2}$, viscous flow of the oxide film or other
inelastic processes can relieve these stresses, in whole or in part. Deal
and Grove \cite{DG} proposed a simple model for the oxidation of planar
substrates in the high-temperature regime, where viscous flow is so rapid
that stress relaxation can be treated as instantaneous. Their model has
become the standard framework within which many subsequent oxidation
problems involving curved geometries \cite{Kao} have been analyzed.

Here we address the opposite regime, where the system is purely elastic and
there is no viscous flow. This case is important for oxidation and other
reactions at lower temperatures. As device structures shrink toward the
nanometer scale, and processing temperatures are correspondingly reduced,
the stress-induced instability we describe may well become important in
systems where it was previously irrelevant.

Consider the geometry of Fig.\ 1, where the substrate-film interface is a
plane modulated by a small sinusoidal perturbation with amplitude $\alpha
_{q}$ and wavelength $\lambda =2\pi /q$. Our task is to evaluate the
interface velocity, and determine whether the deviation from planarity grows
with time. If for some q, $\alpha _{q}$ grows as the reaction proceeds, then
the planar interface is unstable.

The velocity $\vec{v}$ of a point $\vec{x_{I}}$ on the reaction front is
\begin{equation}
\vec{v}(\vec{x_{I}})=R(\vec{x}_{I})\rho _{f}^{-1}\hat{n}_{I}=\left[ c(\vec{%
x_{I}})k_{1}-\rho _{c}k_{2}\right] \rho _{f}^{-1}\hat{n}_{I}~~.
\label{velocity}
\end{equation}
Here $R$ is the number of oxidant molecules per unit area that react at the
interface per unit time to form oxide, $\hat{n}_{I}$ is the local unit
normal to the interface (Fig.\ 1), $\rho _{f}$ is the density of bound
oxidant molecules in the oxide, $c(\vec{x})$ is the concentration of
diffusing oxidant molecules in the film, $k_{1}$ and $k_{2}$ are the
respective rate constants for the forward and reverse reactions (oxidation
and reduction) at the interface, and $\rho _{c}$ is the density of sites
through which oxidant can diffuse in the film. The Fourier component of this
velocity with wave number $q$ gives directly the rate of change of $\alpha
_{q}$, and hence determines the stability of a planar interface.

\bigskip
{\psfig{file=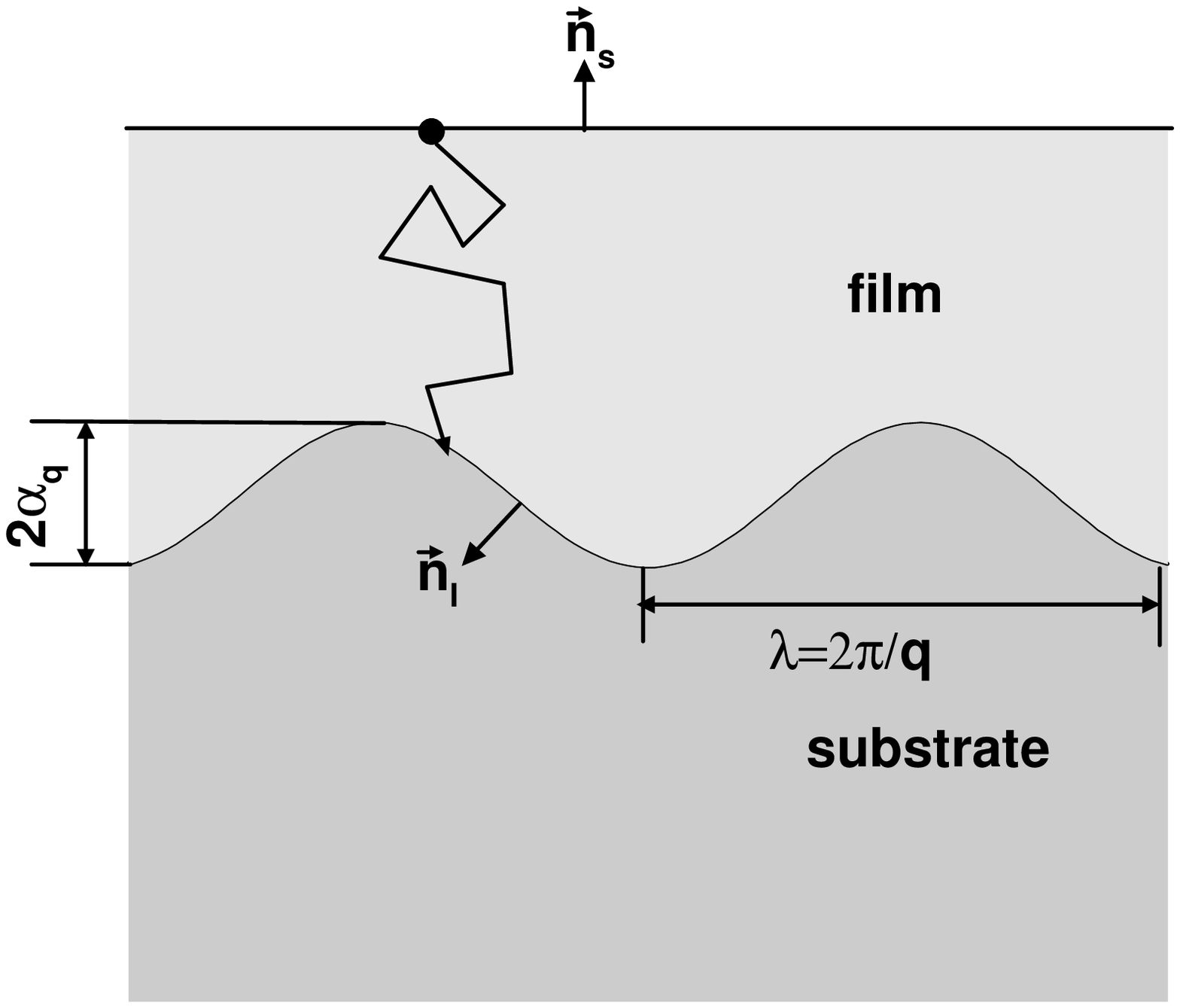,width=200pt}}
{\small FIG.~1. 
An illustration of the geometry of the solid state reaction. The
diffusion of the mobile reactant is represented by a random walk.}
\bigskip

With the current density $\vec{j}$ of diffusing oxidant molecules given by
$
\vec{j} = -D\vec{\nabla}c~~,  \label{current}
$, $c(\vec{x} ,t)$ satisfies the diffusion equation $\partial c /\partial t = D
\nabla^{2} c$, where $D$ is the oxidant diffusion constant. Typically, the
diffusion is rapid enough to maintain a quasi-steady-state concentration
\cite{DG}, so $c$ obeys simply
\begin{equation}
\nabla ^{2}c=0~~.  \label{laplace}
\end{equation}

The boundary condition on this equation at the interface comes from the
requirement that all oxidant molecules flowing into the interface react with
the substrate to produce oxide. This can be written
\begin{equation}
\hat{n}_I \cdot \vec{j}\left( \vec{x_{I}}\right) =R (\vec{x}_I) = c(\vec{%
x_{I}})k_{1}-\rho _{c}k_{2} ~~.  \label{flux3}
\end{equation}
The boundary condition at the upper surface is that the normal oxidant
current equals the rate at which oxidant is incorporated into the upper
surface of the film from the reservoir\cite{DG}:
\begin{equation}
\vec{j}(\vec{x}_{S})=-h\left[ c^{\ast }-c(\vec{x}_{S})\right] \hat{n}_{S}~~.
\label{top/flux}
\end{equation}
Here $h$ is a rate constant (Henry's constant), $\vec{x}_{S}$ is a point on
the upper surface, $c^{\ast }\ \left[ \geq c(\vec{x}_{S})\right] $ is the
concentration of oxidant in the oxide film in equilibrium with the
reservoir, and $\hat{n}_{S}$ is the unit normal to the upper surface.

The solution of Eq.\ (\ref{laplace}) subject to the boundary conditions (\ref
{flux3}) and (\ref{top/flux}) is complicated by the fact that the rate
constants $k_{1}$ and $k_{2}$ depend on the local stress. The stress
dependence has been discussed from a phenomenological perspective by several
authors in the context of silicon oxidation \cite{Kao}. Here we apply a
recent more complete treatment \cite{Tersoff}, which we now briefly sketch.
The reaction proceeds from an initial state of free energy $F_{1}$ to a
final state of free energy $F_{2}$, through a transition state (the saddle
point of the energy surface) with free energy $F_{t}$. The local reaction
rates can then be written as:
\begin{equation}
k_{i}=k_{0}e^{-\beta (F_{t}-F_{i})}~~,  \label{rate}
\end{equation}
where $\beta \equiv 1/k_{B}T$, $k_{B}$ is Boltzmann's constant, $i=1$ or $2$%
, and $k_{0}$ is a rate constant reflecting the microscopic ``attempt
frequency.'' $F_{1}$, $F_{t}$, and $F_{2}$ represent free energies
coarse-grained over distances large compared to atomic dimensions but small
compared to other dimensions of the system. In general, $F_{1}$, $F_{t}$,
and $F_{2}$ depend on the stresses and interface curvature throughout the
system. Note that we are concerned with free-energy {\it changes} $%
F_{i}-F_{t}$ associated with a reaction event, and these depend only on the
{\it local} stress and curvature.

We are considering stability with respect to an infinitesmal perturbation
from planarity, so we expand the stress and reaction rate to lowest order in
$\alpha _{q}$. Then $F_{i}=\bar{F}_{i}+\delta F_{i}$, for $i=1,t,2$. Here $%
\bar{F}_{i}$ is the (local) free energy for the planar interface, including
stress effects, and $\delta F_{i}$ is the change due to the curvature and to
the extra stress produced by the sinusoidal corrugation. Since $\delta F$ is
$O(\alpha _{q}/\lambda )$, we can take $\delta F\ll k_{B}T$, so $%
k_{i}\approx \bar{k}_{i}\left[ 1-\beta (\delta F_{t}-\delta F_{i})\right] $,
with $i=1,2$ and $\bar{k}_{i}=k_{0}e^{-\beta (\bar{F}_{t}-\bar{F}_{i})}$.
The reaction rate is then
\begin{eqnarray}
R(\vec{x}_{I}) &\approx &\bar{R}(\vec{x}_{I})+\beta c(\vec{x}_{I})\bar{k}%
_{1}(\delta F_{t}-\delta F_{1})  \nonumber \\
&&-\beta \rho _{c}\bar{k}_{2}(\delta F_{t}-\delta F_{2})~~,  \label{final}
\end{eqnarray}
where $\bar{R}(\vec{x}_{I})=\left[ c(\vec{x}_{I})\bar{k}_{1}-\rho _{c}\bar{k}%
_{2}\right] $ is the rate for the flat interface.

Expanding $\delta F$ to lowest order in the local strain and curvature gives
\begin{equation}
\delta F_{t}-\delta F_{i}=(\gamma _{i}\kappa +\sigma _{i}^{n}\delta \epsilon
_{nn}+\sigma _{i}^{p}\delta \epsilon _{pp})\rho _{f}^{-1}~~.  \label{free-en}
\end{equation}
Here $\gamma _{i}$ is the curvature-derivative of $F_t - F_i$, and acts as
an effective interface tension; $\kappa $ is the local curvature of the
interface; $\delta \epsilon _{nn}$ and $\delta \epsilon _{pp}$ are the extra
strains normal and parallel to the interface due to deviations from
planarity; and $\sigma _{i}^{n}$ and $\sigma _{i}^{p}$ are coefficients
reflecting the stress of the transition state relative to that of the
initial or final state in the normal and parallel directions respectively
\cite{prev}. Note that the expansion for $\delta F_{t}-\delta F_{i}$ can be
written in terms of the strains $\delta \epsilon _{nn}$ and $\delta \epsilon
_{pp}$ of either the film or the substrate; but the coefficients $\sigma
_{i}^{n}$ and $\sigma _{i}^{p}$ assume different values in the two cases.

We evaluate the strains for the geometry of Fig.\ 1 using linear elastic
theory \cite{LL}. 
The volume expansion accompanying oxidation gives rise to elastic
displacements $\vec{u} (\vec{x} )$ of the material at position $\vec{x}$.
Taking both the substrate and oxide to be isotropic elastic media but with
different elastic constants, one expresses the stress tensor $\stackrel{%
\leftrightarrow}{\sigma}$ in terms of $\vec{u}$ through the standard relation

\begin{eqnarray}
\sigma^{s,f}_{ij} = \lambda^{s,f} (\vec \nabla \cdot \vec u^{s,f} )
\delta_{ij} &+& \mu^{s,f} (\partial u_i^{s,f} / \partial x_j + \partial
u_j^{s,f} /\partial x_i)  \nonumber \\
&-&3K^{s,f} \eta^{s,f} \delta_{ij} ~~.  \label{stress}
\end{eqnarray}
Here $\eta$ is the misfit strain parallel to the surface, with $\eta^s = 0$
and $\eta^f = \eta$ for the substrate (s) and film (f), respectively; $%
K^{s,f}$ are the bulk moduli, $K^{s,f} \equiv \lambda^{s,f} + 2\mu^{s,f} /3$%
; and $(\lambda^{s,f}, \mu^{s,f})$ are the Lam\'{e} coefficients. The above
form of the stress tensor incorporates the requirement that the stress-free
state for the substrate has zero displacement, $\vec u^{s} = 0$, while the
stress-free state for the film is achieved through a uniform stretching or
diagonal strain $\eta$ in all three directions relative to the reference
substrate; i.e., $u_{i}^{f} (\vec{x}) = \eta x_i$ for $i=x,y,z$.

This treatment assumes a fixed misfit at the interface, and would be exact
for an interface between two crystals with a fixed epitaxial relationship,
such as Si-NiSi$_{2}$. However, the case of greatest interest, Si-SiO$_{2}$,
is more complex. The large volume increase upon oxidation is largely
accommodated by expansion normal to the interface, with only a modest
residual misfit stress. Since the microscopic oxidation processes that
determine the residual stress are unknown, our assumption is reasonable but
untested in the context of Si oxidation.

To compute the stresses one must solve the elastic force-balance equation $%
\vec{\nabla} \cdot \stackrel{\leftrightarrow }{\sigma }=0$ for both
substrate and film, subject to the boundary conditions $\sigma \rightarrow 0$
as $z\rightarrow -\infty $ (in the substrate), $\stackrel{\leftrightarrow}{%
\sigma}^{f} \cdot \hat{n}_{S} = 0$ (force balance at the upper surface), $%
\vec{u}^s (\vec{x}_I) = \vec{u}^f (\vec{x}_I)$ (continuity of displacement
at the interface), and $\stackrel{\leftrightarrow}{\sigma}^{s} \cdot \hat{n}%
_{I} = \stackrel{\leftrightarrow}{\sigma}^{f} \cdot \hat{n}_{I} $ (force
balance across the oxide-substrate interface) \cite{IFstress}.

For the planar interface ($\alpha_q = 0$), the displacement and stress must 
vanish in the substrate. For the film, only the z-component of the
displacement is nonzero: $u^{f}_z (z) = \epsilon_{f} z $, with $\epsilon_{f}
= 3 K^{f} \eta / (\lambda^{f} + 2\mu^{f})$. The only nonvanishing components
of $\stackrel{\leftrightarrow}{\sigma}^{f}$ are the diagonal ones parallel
to the interface: $\sigma^{f}_{xx} = \sigma^{f}_{yy} = - 2 \mu^f
\epsilon_{f} $. 

For the planar film, the oxidant current must be uniform throughout the
oxide. So, by solving Eq.\ (\ref{laplace}) with the boundary conditions (\ref
{flux3}), (\ref{top/flux}), one readily calculates the growth rate of the
oxide thickness l(t) from Eq.\ (\ref{velocity}):
\begin{equation}
\frac{d}{dt}l(t)=\frac{(c^{\ast }-\rho_c \bar{k}_{2}/\bar{k}_{1})/\rho_f }{1/%
\bar{k}_{1}+1/h+l(t)/D}~~.
\end{equation}
This result is essentially identical to that of Deal and Grove\cite{DG},
viz., $l(t)$ grows as $t$ at short times and as $t^{1/2}$ at long times.
Note that in practice $c^{\ast }>\bar{k}_{2}\rho _{f}/\bar{k}_{1}$, so $l(t)$
always increases with time.

For the modulated geometry of Fig.\ 1, the interface position is $%
z_{I}(x,t)=z_{0}(t)+\alpha _{q}(t)cos(qx)$, $z_{0}(t)$ being its average
position at time $t$. We calculate the curvature, displacements, and
stresses to linear order in $\alpha _{q}/\lambda $. With Eq.\ (\ref{free-en}%
) one can then express $\delta F_{t}-\delta F_{i}$, and hence the oxidant
flux $\vec{j}(\vec{x}_{I})$ at the interface, in terms of $\alpha _{q}$, $%
\lambda $, the oxide thickness $l$, and the parameters of the problem, such
as $\gamma _{i}$ and $\sigma _{i}^{p,n}$. One then solves $\nabla ^{2}c=0$
in the oxide, to $O(\alpha _{q}/\lambda )$, subject to the boundary
condition (\ref{top/flux}) at the upper surface, and (\ref{flux3}) at the
interface. Knowing $c(\vec{x}_{I})$, one uses Eq.\ (\ref{velocity}) to
determine the interface position $z_{I}(\vec{x},t)$: $z_{0}(t)$ is given by
the solution of the planar problem, and the amplitude, $\alpha _{q}(t)$, of
the perturbation evolves according to:
\begin{equation}
d\alpha _{q} / dt=\Omega _{q}\alpha _{q}~~,  \label{time}
\end{equation}
where
\begin{equation}
\Omega _{q}={\rho _{f}}^{-1}(-w_{0}+w_{1}q-w_{2}q^{2})~~.  \label{w-form}
\end{equation}
The $w_{0}$, $w_{1}$, and $w_{2}$ terms respectively represent the effects
of diffusion, stress, and interfacial tension. Since $w_{0}>0$ and $w_{2}>0$%
, at sufficiently long or short wavelengths $\Omega _{q}$ is always negative
and perturbations decay with time. However, depending on the sign and
magnitude of the stress term $w_{1}$, $\Omega _{q}$ may be positive for some
range of $q$. Perturbations with wavelengths in this range will then grow
exponentially with time, destabilizing the planar interface.

The $w_{i}$ can be written
\begin{eqnarray}
w_{0} &=&\frac{\bar{k}_{1}}{D\rho _{f}}(C_{0}\bar{k}_{1}-r_{c}\bar{k}%
_{2})f_{0}(ql)~~;  \label{w0} \\
w_{1} &=&-\beta \epsilon _{f}[(C_{0}\bar{k}_{1}\sigma _{1}^{n}-r_{c}\bar{k}%
_{2}\sigma _{2}^{n})f_{1}^{n}(ql)+  \nonumber \\
&&(C_{0}\bar{k}_{1}\sigma _{1}^{p}-r_{c}\bar{k}_{2}\sigma
_{2}^{p})f_{1}^{p}(ql)]~~;  \label{w1} \\
w_{2} &=&\beta (C_{0}\bar{k}_{1}\gamma _{1}+r_{c}\bar{k}_{2}\gamma
_{2})f_{2}(ql)~~.  \label{w2}
\end{eqnarray}
Here
\[
C_{0}\equiv \frac{c^{\ast }\rho _{f}^{-1}+r_{c}\bar{k}_{2}/h+r_{c}\bar{k}%
_{2}l/D}{(1+\bar{k}_{1}/h+\bar{k}_{1}l/D)}\ ,
\]
and $r_{c}\equiv \rho _{c}/\rho _{f}$ is of order unity. The quantities $%
f_{0}(ql)$, $f_{1}^{n}(ql)$, $f_{1}^{p}(ql)$, and $f_{2}(ql)$ are
complicated dimensionless functions of $ql$ and the various parameters. The
dependence of these functions on $ql$ is weak and they are of
order unity over the range of $ql$'s of interest. 
For many
reactions, including the oxidation of silicon, $\bar{k}_{2}$ is negligibly
small. Thus the reverse reaction (decomposition of the oxide) doesn't occur
in practice. In this case, formulae (\ref{w0}-\ref{w2}) simplify, since $%
\bar{k}_{2}$ can be set to 0.

The explicit $q$-dependence of the last two terms in Eq.\ (\ref{w-form})
results from the strains ($\delta \epsilon _{nn}$ and $\delta \epsilon _{pp}$%
) and the curvature ($\kappa $) behaving like $q$ and $q^{2}$, respectively.
Thus the smoothing influence of the oxidant diffusion term dominates and
maintains stability at small $q$. This is the inverse of diffusion-limited
{\it growth}, where the role of diffusion is destabilizing, as in the
Mullins-Sekerka instability. The interface tension term prevents
perturbations of very small wavelength from growing, stabilizing the planar
interface at large $q$.

Thus planar growth is unstable when $\Delta=w_{1}^2-4w_{0}w_{2}>0$ with the
instability $\Omega (q)>0$ occurring only over the range of wavelengths $%
q_{-}<q<q_{+}$ . Here
\[
q_{\pm }=(w_{1}\pm \sqrt{w_{1}^{2}-4w_{0}w_{2}} ~~)/2w_{2}
\]
This is illustrated in Fig.\ 2.

\bigskip
{\psfig{file=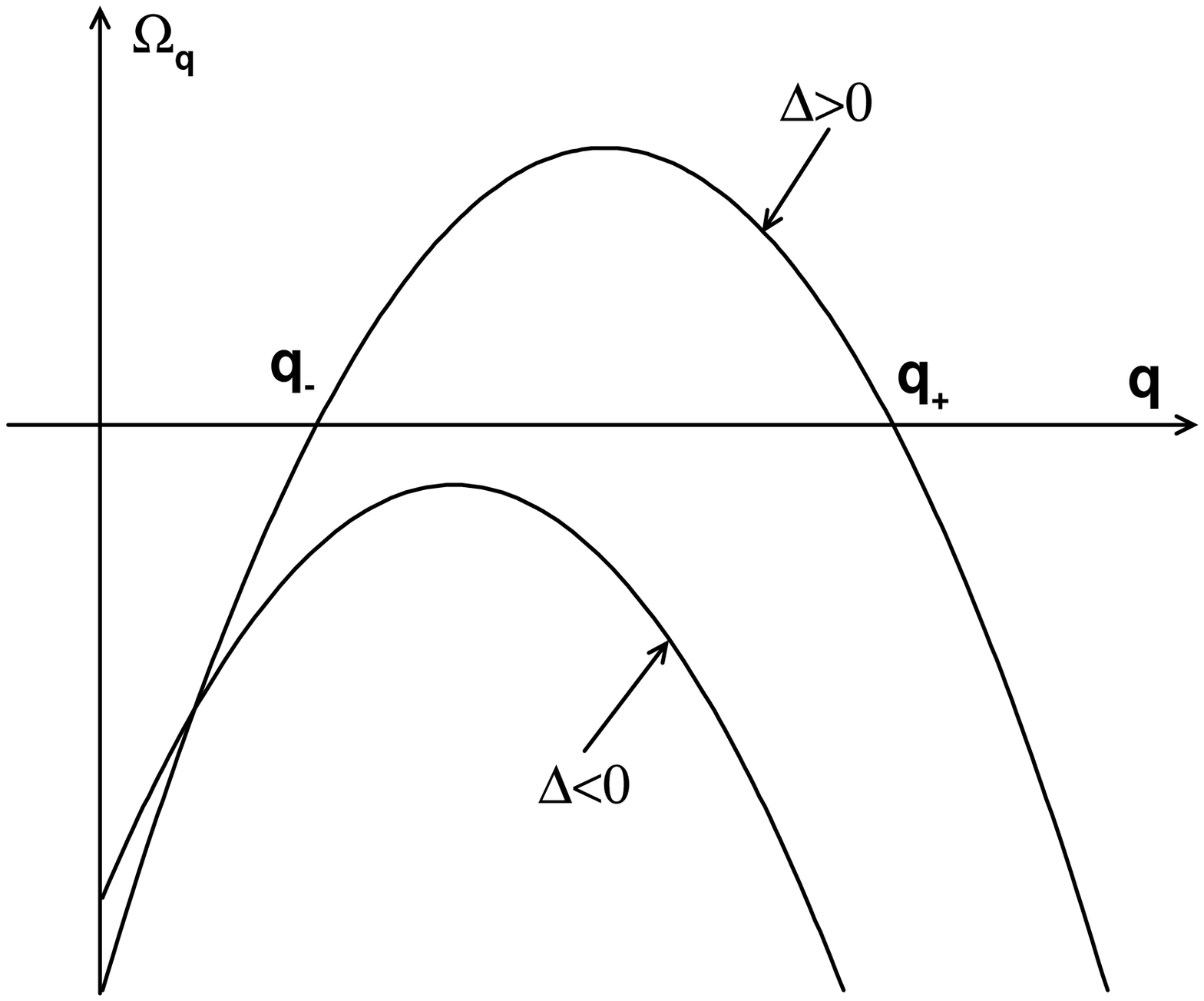,width=200pt}}
{\small FIG.~2. 
An illustration of the stability calculation results. Notice that
the interface is always stable ($\Omega_{q}<0$) in both the small and large $%
q$ limit, and only become unstable ($\Omega_{q}>0$) in the intermediate
range $q_{-}<q<q_{+}$ when $\Delta>0$.}
\bigskip

It remains to determine the typical range of wavenumbers $q$ for which the
instability holds. We first simplify the problem by assuming that $\bar{k}%
_{2}$ is indeed negligibly small. As mentioned above, this assumption is
typically well justified in practice. We must also estimate the parameters $%
\sigma _{1}^{n}$ and $\sigma _{1}^{p}$. Because oxidation is accompanied by
expansion, the oxidation reaction should proceed more readily when the
existing material near the interface has been dilated, i.e., when $\delta
\epsilon _{nn}$ and $\delta \epsilon _{pp}$ are positive. In other words,
one expects both coefficients $\sigma _{1}^{n}$ and $\sigma _{1}^{p}$ to be
negative. Our elastic theory calculation shows, however, that $%
f_{1}^{n}(ql)<0$ and $f_{1}^{p}(ql)>0$ (with $f_{1}^{n}(ql)+f_{1}^{p}(ql)>0$%
), so the sign of $w_{1}$ is determined by the relative magnitudes of $%
\sigma _{1}^{n}$ and $\sigma _{1}^{p}$. Thus the instability presumably will
not occur for all values of parameters.

To see that it can nonetheless occur for some reasonable parameters, we
estimate $\sigma _{1}^{n}$ and $\sigma _{1}^{p}$ by assuming $\sigma
_{1}^{n}\sim \sigma _{1}^{p}\sim -2\mu _{f}\epsilon ^{\prime }$, which makes
$w_{1}>0$. For the other parameters, we use rough numbers for wet oxidation
in silicon\cite{DG}. Due to the large volume change during oxidation, the
``effective'' strain parameter $\epsilon ^{\prime }$ is of order unity: $%
\epsilon ^{\prime }\sim 1$. Taking $l(t)\ll D/\bar{k}_{1}$ (the most
favorable limit for the occurrence of the instability), $c^{\ast }\sim 3$x$%
10^{19}~cm^{-3}$, $\bar{k}_{1}\sim 5$x$10^{-5}cm/sec$, $D\sim
10^{-9}cm^{2}/sec$, $\rho _{f}\sim 10^{22}cm^{-3}$, $w_{2}\sim 10^{9}/cm-sec$%
, and $\beta \mu ^{f}\sim 1.6$x$10^{24}cm^{-3}$ at about $900^{o}C$, we find
that the inequality $w_{1}>2\sqrt{w_{0}w_{2}}$, and hence the instability,
holds for all $\epsilon _{f}$ greater than about $2\times 10^{-4}$, which
can indeed be achieved for modest values of misfit strain $\eta $. For the
numbers above, we find, roughly, $q_{-}\sim 2$x$10^{4}/cm$, and $q_{+}\sim 5$%
x$10^{7}/cm$ for $\epsilon _{f}=0.01$. This implies an instability for
wavelengths between roughly 1 $nm$ and 0.3 $\mu m$.

Eq.\ (\ref{time}) shows that the characteristic time $\tau _{I}$ over which
the instability develops is set by $1/\Omega _{q_{max}}=4w_{2}\rho
_{f}/(w_{1}^{2}-4w_{0}w_{1})$, where $q_{max}\equiv (q_{+}+q_{-})/2$ is the
wavenumber for which $\alpha _{q}$ grows most rapidly. For the sample
parameters above, one finds that $\tau _{I}\sim 1~sec$. This time scale is
short compared with typical processing times, or with the other
characteristic time $\tau _{X}=D\rho _{f}/2c^{\ast }{\bar{k}_{1}^{2}}\sim
10^{4}sec$, which is roughly the time when the motion of the planar
interface crosses over from $t$ to $t^{1/2}$ behavior.

Finally, we note that there is some intriguing experimental evidence of
roughness in $Si-SiO_{2}$ interfaces for thin oxide layers \cite{Sinclair}
at 900$^{o}C$. However, because of the poorly-understood complexities of
real Si oxidation, with a volume expansion of order unity but a far smaller
residual misfit, our results should be viewed as indicative of the kind of
behavior that may occur, rather than as quantitatively applicable.



\end{multicols}


\begin{references}
\bibitem{prior}  Related stress-induced instabilities at reaction fronts
have been independently analysed in Refs.\ \cite{Freund} and \cite{Aziz}.
Ref.\ \cite{Freund} is a preliminary report with limited detail, but
apparently uses a very different model for the effect of stress
on the reaction rate.
Ref.\ \cite{Aziz} focuses on the
stress dependence of the reaction rate,
analysing a special case where no diffusion is required for the reaction.
\bibitem{prev}  Previous treatments of the stress dependence of the reaction
rate (ref. \cite{Kao}), have neglected the transverse component of stress or
strain, and focussed on the volume expansion of the final oxide state,
rather than on the intermediate state.
\bibitem{IFstress}  In principle there is also an interface stress, which
contributes a force at a nonplanar interface; but this term is typically
negligible, and is therefore omitted here.
\end{references}
\end{document}